\documentclass[]{jfm}

\usepackage{graphicx,}
\usepackage[justification=justified]{caption}
\usepackage{newtxtext}
\usepackage{newtxmath}
\usepackage{natbib}
\usepackage{hyperref}
\usepackage{verbatim}
\usepackage{xcolor}
\usepackage{marginnote}
\hypersetup{
    colorlinks = true,
    urlcolor   = blue,
    citecolor  = black,
}
\newtheorem{lemma}{Lemma}
\newtheorem{corollary}{Corollary}
\newcommand{\RomanNumeralCaps}[1]
\linenumbers

\newcommand\mc[1]{{\color{black}#1}}
\newcommand\mcc[1]{{\color{black}#1}}

\title{
Mixing across stable density interfaces in forced stratified turbulence
}
\shorttitle{Mixing across stable density interfaces}

\author{Miles M. P. Couchman\aff{1}
  \corresp{\email{mc2277@cam.ac.uk}},
  Stephen M. de Bruyn Kops\aff{2}
 \and Colm-cille P. Caulfield\aff{1,3}}

\affiliation{\aff{1}Department of Applied Mathematics and Theoretical Physics, University of Cambridge, Cambridge CB3 0WA, UK
\aff{2}Department of Mechanical and Industrial Engineering, University of Massachusetts Amherst, Amherst, MA 01003, USA
\aff{3}Institute for Energy and Environmental Flows, University of Cambridge, Cambridge CB3 0EZ, UK}

\begin{document}
\maketitle

\begin{abstract}
Understanding how turbulence enhances irreversible scalar mixing in density-stratified fluids is a central problem in geophysical fluid dynamics. \mcc{While isotropic overturning regions are commonly the focus of mixing analyses, we here investigate whether significant mixing may arise in anisotropic statically-stable regions of the flow}. \mcc{Focusing on a single} forced direct numerical simulation of stratified turbulence, \mcc{we analyze} spatial correlations between the vertical density gradient $\partial\rho/\partial z$ and the dissipation rates of kinetic energy $\epsilon$ and scalar variance $\chi$, the latter quantifying scalar mixing. \mcc{The domain is characterized by relatively well-mixed density layers separated by sharp stable interfaces that are correlated with high vertical shear.} \mcc{While static instability is most prevalent within the mixed layers}, much of the scalar mixing is localized to the intervening interfaces, a phenomenon not apparent if considering local static instability or $\epsilon$ alone. While the majority of the domain is characterized by the canonical flux coefficient $\Gamma\equiv\chi/\epsilon=0.2$, often assumed in ocean mixing parameterizations, extreme values of $\chi$ within the statically-stable interfaces, associated with elevated $\Gamma$, strongly skew the bulk statistics. Our findings suggest that current parameterizations of turbulent mixing may be biased by undersampling, such that the most common, but not necessarily the most significant, mixing events are overweighted. \mcc{Having focused here on a single simulation of stratified turbulence, it is hoped that our results motivate a broader investigation into the role played by stable density interfaces in mixing, across a wider range of parameters and forcing schemes representative of ocean turbulence.}
\end{abstract}



\section{Introduction}
In a density-stratified fluid, turbulence enhances the rate at which scalars are irreversibly diffused throughout the flow, a process of great interest in a variety of geophysical, environmental and industrial settings (e.g. \cite{fernando1991turbulent}). Of particular importance is characterizing the role of turbulence in the vertical transport of heat within the ocean, a crucial mechanism for driving the required upwelling of cold bottom waters to maintain the ocean's vertical stratification profile and to complete global circulation currents \citep{wunsch2004vertical}. Turbulence in the ocean generates dynamically relevant motions on the order of millimeters, which cannot currently be resolved in numerical circulation models and must therefore be parameterized, with the choice of parameterization found to influence future climate projections  strongly \citep{whalen2020internal}. Considerable observational, numerical and theoretical work has thus been focused on developing more accurate and universal mixing models which account for the wide range of turbulent processes observed in different flow regimes within the ocean \citep{colm2020open}.

The rate at which turbulence mixes a non-uniform density field is often defined in terms of an appropriately-averaged vertical density flux $B \equiv \left\langle \rho'w'\right\rangle $, where $\rho'$ and $w'$ denote fluctuations in density and vertical velocity away from the mean flow, respectively. \mcc{If $B$ is to be used as a robust indicator of irreversible mixing, it is critical that measurements of $B$ are averaged over sufficiently large spatial volumes or time intervals, in order to isolate irreversible diffusive processes from  reversible stirring  motions \citep{villermaux2019mixing}.} Stirring, occurring on relatively large scales, may be thought of as the adiabatic rearrangement of fluid parcels of different density induced by the underlying turbulence, which in principle is reversible. \mcc{Hence, a pointwise measurement of $B$ would not be a sufficient indicator that irreversible mixing had occurred, as the sign of $B$ could subsequently switch direction yielding a net flux of zero.} Thus, we here use the term mixing to refer specifically to the diffusive transport of density across gradients that have been enhanced by such macroscopic stirring motions, irreversibly leading the system toward a state of greater homogenization. In order to isolate only irreversible contributions to mixing, \citet{lorenz1955available} introduced the concept of an available potential energy (APE). APE quantifies the difference between a system's current potential energy and its minimum background potential energy (BPE) that could be achieved if fluid parcels were adiabatically sorted into their most stable configuration. For a Boussinesq fluid, \citet{winters1995available} demonstrated that irreversible mixing may be described as the conversion of APE into BPE, with a system's BPE increasing in time as it homogenizes. Generalizing this mixing framework to compressible flows, \citet{tailleux2009energetics} argued that the mixing \mc{of a thermally-stratified fluid} should most rigorously be defined as the conversion of APE into internal energy, which in the Boussinesq limit then exactly matches the generation of BPE.

\mcc{Given a variety of sampling limitations involved with collecting turbulence data within the ocean, it is exceedingly difficult to perform the averaging required to extract the irreversible component of density fluxes from direct observational measurements of $B$.} Therefore, a number of indirect methods have been proposed that infer such fluxes from more readily available quantities\mcc{, which may be computed locally} \citep{gregg2018mixing}. Two such quantities, associated with what is conventionally referred to as turbulent microstructure, are the dissipation rates of kinetic energy $\epsilon$ and scalar variance $\chi$,
\begin{equation}
\epsilon=\frac{\nu}{2}\left(\frac{\partial u'_{i}}{\partial x_{j}}+\frac{\partial u'_{j}}{\partial x_{i}}\right)^{2}, \quad \chi=\frac{g^{2}\kappa}{\rho_{0}^{2}N^{2}}\left(\frac{\partial\rho'}{\partial x_{i}}\frac{\partial\rho'}{\partial x_{i}}\right), \label{eq1}
\end{equation}
representing the rates at which viscosity $\nu$ and molecular diffusivity $\kappa$ smooth gradients in the turbulent velocity $\mathbf{u'}$ and density $\rho'$ fields, respectively. In equation (\ref{eq1}), $g$ denotes the gravitational acceleration, $\rho_0$ a reference background density and $N=\sqrt{\left(-g/\rho_{0}\right)\partial\overline{\rho}/\partial z}$ 
the buoyancy frequency, defined by an appropriately averaged ambient density gradient $\partial \overline{\rho} / \partial z$ \mc{against which the turbulence acts}. The quantities $\epsilon$ and $\chi$ are intimately related to the irreversible processes associated with the conversion of kinetic energy and available potential energy into internal energy, respectively, as is further described by \cite{caulfield2021layering}. \mc{In particular, for the class of direct numerical simulation considered here, characterized by an imposed uniform background stratification $N^2_0$, \cite{howland2021quantifying} demonstrated that $\chi$ computed using $N^2=N_0^2$ in equation (\ref{eq1}) provides an excellent approximation to the destruction rate of APE and is therefore a \mcc{good} measure of local irreversible mixing.} 

As discussed by \cite{ivey2018quantifying}, $\chi$ also arguably provides the most robust method for estimating irreversible mixing from oceanographic measurements, 
since $\left\langle B\right\rangle  \simeq \left\langle \chi\right\rangle $ in steady-state provided that averaging is performed over sufficiently long times and large volumes so that 
reversible processes and transport terms are negligible \citep{osborn1972oceanic}. \mcc{Importantly, $\chi$ is both directly proportional to the scalar diffusivity $\kappa$ and sign-definite, providing a robust \textit{local} measure of the irreversible fluxes associated with molecular diffusion, which does not require the averaging of the density flux $B$ needed to filter our reversible local stirring motions in the turbulent flow}. Due to a scarcity of $\chi$ measurements, however, $\epsilon$ is more commonly used to infer mixing following the method of \cite{osborn1980estimates}, which requires the introduction of a flux coefficient $\Gamma \equiv \chi / \epsilon$ to prescribe the fraction of turbulent kinetic energy that leads to irreversible mixing, as opposed to being directly dissipated by viscosity. A constant value $\Gamma = 0.2$ is commonly assumed when estimating global patterns of oceanic mixing \citep{mackinnon2017climate}, \mc{which has been found to be in agreement with tracer release experiments \citep{gregg2018mixing}. However, there is significant evidence suggesting that $\Gamma$ varies appreciably in different flow regimes \citep{caulfield2021layering} and so a clear physical picture has not yet emerged explaining why $\Gamma=0.2$ is a reasonable assumption.}

In the absence of measurements of $\epsilon$ or $\chi$, mixing locations are primarily inferred from the presence of unstable overturns in vertical density profiles, as proposed by \cite{thorpe1977turbulence}.  Assuming that the vertical extent of an overturn is correlated with the Ozmidov length $L_{O}=\sqrt{\epsilon/N^{3}}$, $\epsilon$ may be inferred from the measurement of overturns which can then be converted into a flux via $\Gamma$. However, this assumed correlation between the vertical overturning scale and $L_O$ is not always robust, as has recently been discussed, for example, by \cite{ivey2018quantifying}, \cite{ijichi2020variable} and \cite{mashayeketal21}. Using a forced direct numerical simulation (DNS) similar to that considered here, \cite{taylor2019test} quantified the errors associated with the indirect flux estimates of \cite{osborn1972oceanic}, \cite{osborn1980estimates} and \cite{thorpe1977turbulence} by sparsely sampling vertical profiles of the computational domain in order to mimic oceanographic measurements. 

Spatio-temporal intermittency in stratified turbulence greatly reduces the applicability of classical turbulence modeling assumptions, including the common assumption of log-normal distributions for $\epsilon$ and $\chi$ \citep{debk15}. \cite{cael2021log} found that global ocean measurements of $\epsilon$ were not well approximated by an assumed log-normal distribution but instead had a skewed right tail, indicating that a small number of extreme events dominated the bulk statistics. By considering local correlations between direct ocean measurements of $\epsilon$ and $\chi$, \cite{couchman2021data} further emphasized the importance of extreme events, finding that while the majority of the sampled domain 
was characterized by the canonical flux coefficient $\Gamma = 0.2$, isolated mixing events containing the largest $\chi$ were not reflected by a corresponding local increase in $\epsilon$, yielding a dramatic increase in $\Gamma$.

Vertical layering is also known to be a canonical feature of stratified turbulent flows, with the density field often forming `staircases' of deep, relatively well-mixed layers separated by thin interfaces with strong gradients \citep{caulfield2021layering}. For sufficiently stratified environments, vertical shearing induced by the decoupling of horizontal and vertical motions in such a layered structure becomes an important mechanism for triggering instability and the ensuing generation of turbulence \citep{lilly1983stratified,billant2001self}. Parameterizations of mixing based on simple domain averages are thus unlikely to be accurate as rare extreme events and spatial heterogeneity within the flow will be missed, a potential cause of the highly-scattered mixing statistics currently reported throughout the literature \citep{gregg2018mixing}. 

In an attempt to classify such intermittency in an automatic, yet robust and interpretable manner, \cite{portwood2016robust} devised an algorithm for splitting a snapshot from a forced DNS into three dynamically distinct regions: quiescent regions, intermittent layers and turbulent patches. These regions were distinguished by an increasing degree of local overturning, as determined by computing the fraction of unstable points $\partial \rho / \partial z > 0$ within an extended neighbourhood. Local overturning fractions and dissipation rates $\epsilon$ were found to be strongly correlated, in agreement with the arguments of \cite{thorpe1977turbulence}. For the relatively large filter sizes used to segment the domain, on the order of a buoyancy length $L_B = 2\pi u_h /N$ (where $u_h$ denotes a characteristic horizontal velocity scale), distributions of $\chi$ associated with each region were also found to be correlated with $\epsilon$, although the finer spatial distributions of $\epsilon$ and $\chi$ within each region, and the resulting flux coefficient $\Gamma$, were not considered.

Motivated by the automated flow segmentation of \cite{portwood2016robust} in terms of unstable local density gradients $\partial \rho / \partial z > 0$, and the observation of \cite{couchman2021data} that, within the ocean, extreme events in $\chi$ are not necessarily correlated with those in $\epsilon$, we here analyze spatial mixing distributions within a computational domain by considering local correlations between $\epsilon$, $\chi$ and $\partial \rho / \partial z$. In particular, we wish to probe whether overturning alone provides a robust indicator for local mixing, or if significant mixing as revealed by $\chi$ might occur in other regions that would seem inconspicuous based on consideration of only $\epsilon$ or $\partial \rho / \partial z$. 

In line with the previous investigations of \cite{portwood2016robust} and \cite{taylor2019test}, we consider a forced DNS of stratified turbulence using the methodologies presented in \cite{almalkie2012kinetic}. In $\S$\ref{sec:2_Dataset}, we summarize the DNS dataset considered here and highlight the presence of a (previously-unreported) robust vertically-aligned vortex generated by the forcing scheme, that injects energy into the domain at large scales and induces vertical layering in the surrounding flow. In $\S$\ref{sec:3_PointwiseStats}, we then consider pointwise correlations between $\partial \rho / \partial z$, $\epsilon$, $\chi$ and the flux coefficient $\Gamma$, which suggests that mixing occurs not only in overturning regions, but also in areas of local static stability. In $\S$\ref{sec:4_MixingStructures}, we move beyond pointwise statistics to consider extended mixing structures within the flow, highlighting two ways in which local static instability in the density gradient fails to be a sufficient indicator of mixing: within the vortex a lateral density gradient is correlated with the majority of $\chi$, and outside the vortex extreme values of $\chi$ are localized to relatively `sharp' stable density interfaces at the bounding edges between overturning layers. Finally, in $\S$\ref{sec:5_Discussion}, we summarize our results and discuss implications for parameterizing turbulent mixing within the ocean. 

\section{Summary of DNS dataset} \label{sec:2_Dataset}
We consider a statistically-steady, forced DNS of stratified turbulence from the simulation campaign originally reported by \cite{almalkie2012kinetic}, and subsequently analyzed by \cite{portwood2016robust} and \cite{taylor2019test}. Using a characteristic root-mean-square horizontal velocity scale $u_h$, length scale $L$, and background buoyancy frequency $N_0$, the non-hydrostatic Boussinesq approximation of the Navier-Stokes equations may be written in the following dimensionless form:

\refstepcounter{equation}
$$
\nabla\cdot\mathbf{u}=0, \quad
\frac{\partial\mathbf{u}}{\partial t}+\mathbf{u}\cdot\nabla\mathbf{u}=-\left(\frac{2\pi}{Fr}\right)^{2}\rho\hat{\mathbf{z}}-\nabla p+\frac{\nabla^{2}\mathbf{u}}{Re}+\mathcal{F}, \quad
\frac{\partial\rho}{\partial t}+\mathbf{u}\cdot\nabla\rho-w=\frac{\nabla^{2}\rho}{RePr}.
\eqno{(\theequation{\mathit{a},\mathit{b},\mathit{c}})}\label{eq35}
$$

The governing equations (\ref{eq35}) are numerically integrated using a pseudospectral technique in a triply-periodic domain, as detailed by \cite{almalkie2012kinetic}. The dimensionless parameters governing the flow are the Prandtl number $Pr = \nu/ \kappa$, Froude number $Fr_h = 2 \pi u_h / (N_0 L)$ and Reynolds number $Re_h = u_h L / \nu$. \mcc{The density field satisfies}
\begin{equation}
\mcc{\rho\left(\boldsymbol{x},t\right)=\rho_{0}(1 -N_{0}^{2}z/g)+\rho'\left(\boldsymbol{x},t\right)}, \label{eqRho}
\end{equation}
\mcc{where $\rho_{0}(1 -N_{0}^{2}z/g)$ defines a time-independent, linear background density gradient characterized by a reference density $\rho_0$ and an imposed constant background buoyancy frequency $N_0$. Density perturbations $\rho'$ away from this linear background state satisfy the periodic boundary conditions and are used to compute $\chi$ in equation (\ref{eq1}).} The imposed constant background buoyancy frequency $N_0$ is used as the characteristic `appropriately-averaged' buoyancy frequency $N$ required to compute $\chi$ in equation (\ref{eq1}), \mc{as is widely considered the natural choice when quantifying irreversible mixing in numerical simulations with an imposed background stratification (see e.g. \citet{shih2005parameterization, maffioli2016mixing, garanaik2019inference, portwood2019asymptotic}). By explicitly computing the available potential energy (APE) of a triply-periodic domain with an imposed uniform background stratification $N_0$, \citet{howland2021quantifying} confirmed that normalizing $\chi$ by $N_0$ indeed provides an excellent approximation to the true irreversible mixing rate as computed through changes in the system's APE.} The forcing term $\mathcal{F}$ is governed by the deterministic scheme denoted `Rf' in \citet{rao2011mathematical}, which adds energy to horizontal motions larger than 1/8th of the horizontal box size so as to match a target kinetic energy spectrum at small wavenumbers. We consider a simulation characterized by $Pr = 1$, $Fr_h = 2.23$ and $Re_h = 1271$, in a domain of size $2 \pi \times 2 \pi \times \pi$ with $4096 \times 4096 \times 2048$ grid points, resulting in a grid spacing of $\Delta \approx L_K/2$, with $L_K$ denoting the Kolmogorov length scale. \mc{For reference, the characteristic buoyancy Reynolds number of the simulation is $Re_{b}=\left\langle \epsilon\right\rangle /\nu N_{0}^{2}=50$.} We consider a single snapshot of the flow in time and all figures are displayed on grids that have been sparsed by a factor of eight in each dimension. 

\begin{figure}
	\captionsetup{width=1\linewidth}
	\centerline{\includegraphics[width=1\textwidth]{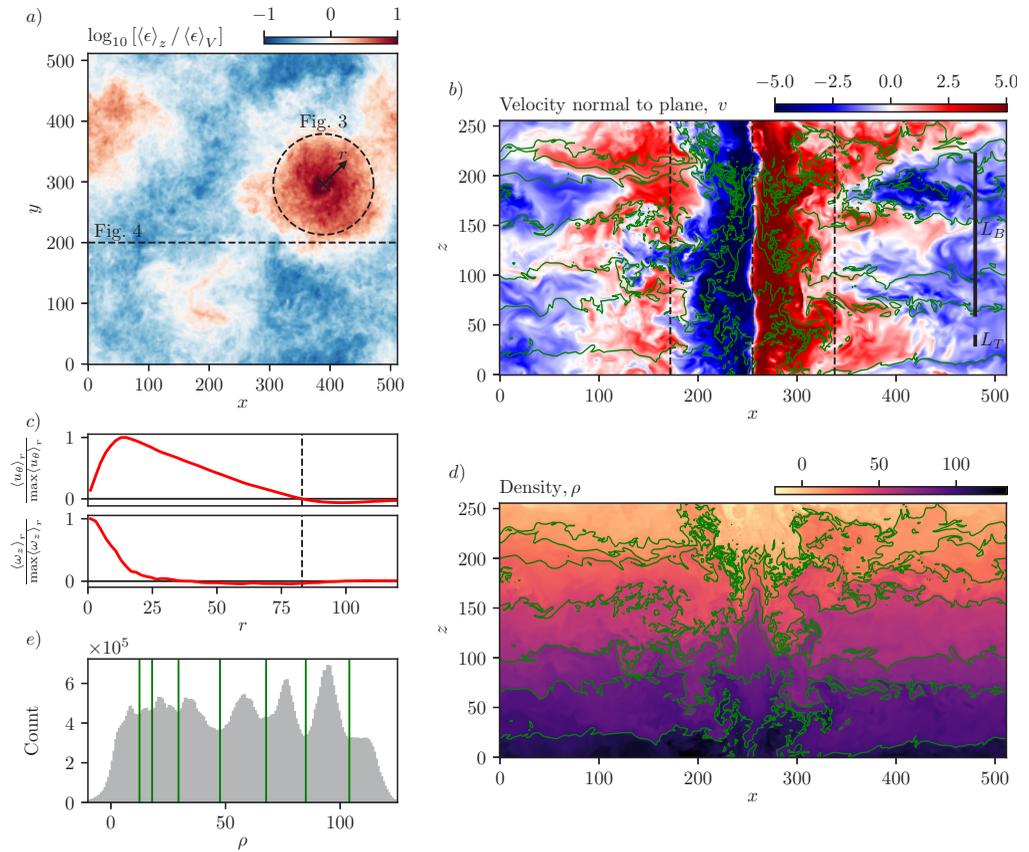}}
	\caption{Large-scale characteristics of the velocity and density fields within the computational domain. a) The vertical average of the dissipation rate $\epsilon$ at gridpoint $(x,y)$, normalized by the domain average. The whole domain is shown, with gridpoints sparsed by a factor of eight in each dimension. The dashed circle highlights a region of elevated $\epsilon$, coinciding with a vortex in the velocity field, \mc{as is further examined in Figure \ref{fig:Fig3}. A vertical slice of the domain at $y=200$ is considered in Figure \ref{fig:Fig4}.} b) The horizontal velocity normal to the plane for a vertical slice at constant $y$ passing through the center of the dashed circle in panel a), revealing a vertically-aligned vortex rotating counterclockwise. The grid has been shifted in $x$ relative to panel a) so as to center the vortex. The buoyancy length $L_B$ and Taylor length $L_T \approx 25 L_K$ (where $L_K$ is the Kolmogorov length) are marked for reference, as were the filter sizes used in the segmentation analysis of \cite{portwood2016robust}. Green contours mark stable interfaces in the density field, as shown in panel d). c) The radially-averaged angular velocity $u_\theta$ and vertical component of vorticity $\omega_z$ as a function of the distance $r$ from the center of the vortex. The dashed line at $r\approx 80$ corresponds to the dashed circle plotted in panel a). d) The density field corresponding to the vertical slice of velocity plotted in panel b). The green lines in panels b) and d) illustrate contours at the minimum values of the histogram of the density field in panel e), as are marked with vertical green lines, delineating the interfaces between relatively well-mixed density layers in the flow surrounding the vortex.}
	\label{fig:Fig1}
\end{figure}

The main characteristics of the dataset are summarized in Figure \ref{fig:Fig1}. The vertically-averaged dissipation rate $\epsilon$ (Figure \ref{fig:Fig1}a) reveals a dominant patch of elevated turbulence that is generated by a large-scale vertically-aligned vortex in the velocity field, rotating counter-clockwise (Figure \ref{fig:Fig1}b). Radial averages, centered on the vortex, of the angular velocity $u_\theta$ and vertical component of vorticity $\omega_z = \partial v/\partial x-\partial u/\partial y$ are plotted in Figure \ref{fig:Fig1}c, indicating a Rankine-type vortex that is approximately characterized by rigid body rotation at small radii $r$ from the vortex core, followed by a transition to roughly irrotational flow at larger $r$. It is important to note that such a description characterizes the radially-averaged flow, and that smaller-scale vortical motions will still certainly be present in the turbulent patch surrounding the vortex. A series of horizontal currents traveling in alternating directions are found to emanate from the vortex, characterized by a vertical scale on the order of a buoyancy length $L_B$. In Figure \ref{fig:Fig1}d, we plot the vertical slice of the density field that corresponds to the velocity field shown in Figure \ref{fig:Fig1}b, highlighting an analogous vertically-layered structure outside of the vortex, with relatively well-mixed density layers separated by sharp, stable interfaces. The approximate locations of these density interfaces (delineated by green contours) correspond to minima in the histogram of $\rho$ (Figure \ref{fig:Fig1}e), which highlights a strong perturbation of the density field away from its uniform background gradient. Superimposing these density contours on the velocity field in Figure \ref{fig:Fig1}b highlights that the sheared interfaces in the velocity field are strongly correlated with the stable interfaces in the density field. This correlation is further demonstrated in Supplementary Video 1, where rotations of the slices in Figures \ref{fig:Fig1}b,d around the center of the vortex are shown. In $\S$\ref{sec:4_MixingStructures}, we demonstrate that these interfaces, characterized by both high shear and a strong statically-stable density gradient, are critically important for the mixing generated outside of the vortex. 

The spontaneous formation of a persistent vortex is a key, yet previously unreported feature of the forcing scheme of \citet{rao2011mathematical} used to generate statistically-steady turbulence. In particular, the identification of the vortex in Figure \ref{fig:Fig1} provides insight into how the segmentation results of \citet{portwood2016robust} (see their Figure 2c), who used an identical forcing scheme, are related to the background flow field. Specifically, the roughly cylindrical patch of most vigorous turbulence detected by \citet{portwood2016robust}, using a filter of size $L_B$, extends across the entire vertical domain and almost certainly corresponds to an analogous vortical structure in their DNS. Similarly, their `intermittent layers' are primarily composed of horizontal offshoots from the central vertically-aligned turbulent patch, and are shaped by a similar pattern to the sheared velocity interfaces observed in Figure \ref{fig:Fig1}b. As it is now evident that \citet{portwood2016robust} have broadly identified such a vortex to be a turbulent hotspot, a goal of this study is to perform a finer analysis of mixing patterns both within and outside of the vortex, in order to determine how patterns in the small-scale turbulent microstructure, as described by $\epsilon$ and $\chi$, are related to the larger-scale layered structure of the flow.

\section{Pointwise statistics conditioned on local density gradient} \label{sec:3_PointwiseStats}
\begin{figure}
	\captionsetup{width=1\linewidth}
	\centerline{\includegraphics[width=1\textwidth]{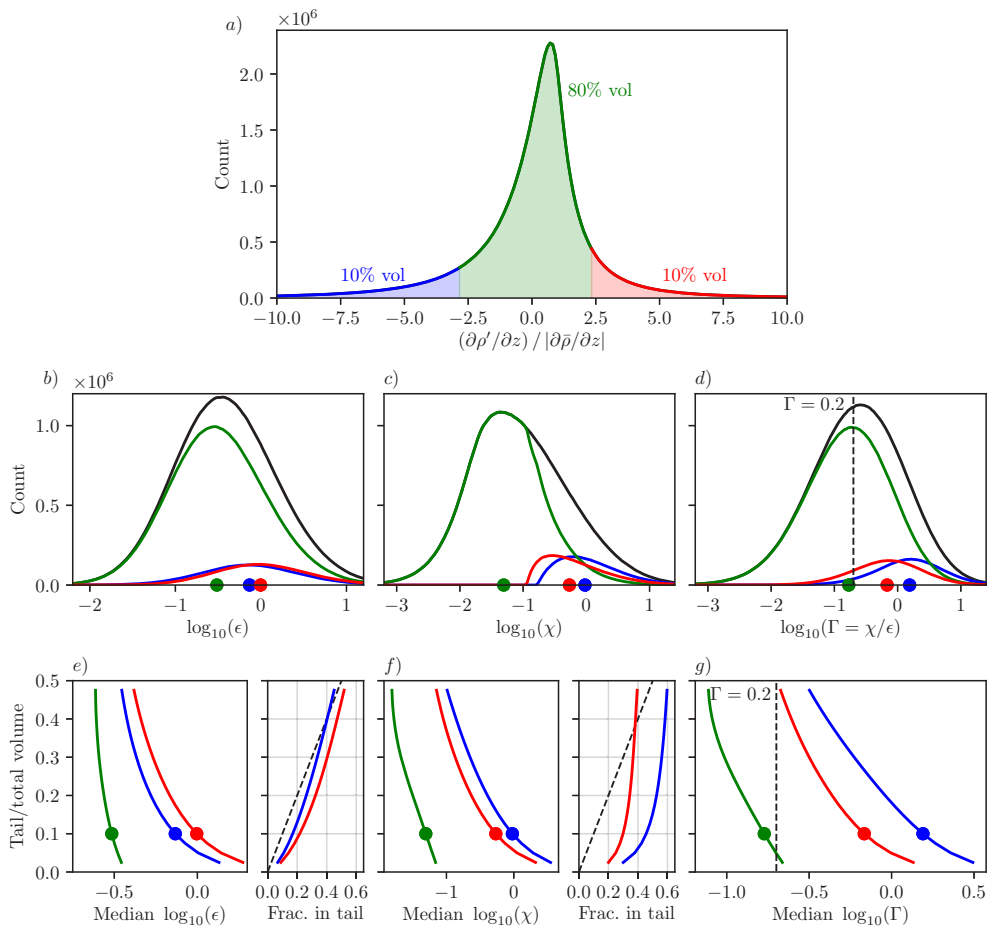}}
	\caption{Pointwise mixing statistics of the DNS data, conditioned by the local vertical density gradient. a) Histogram of the perturbed density gradient $\partial \rho' / \partial z$ normalized by the magnitude of the imposed uniform background gradient, with values greater than one indicating local overturning. The distribution is split into three regions, by assigning a fixed volume (here $10\%$) to each tail. Panels b)-d) illustrate the distributions of $\epsilon$, $\chi$ and $\Gamma$, respectively, for the whole domain (black) and the subdomains encompassed by the coloured regions in panel a). Circles mark the median values of each distribution, and the dashed line in panel d) indicates the canonical flux coefficient $\Gamma = 0.2$ for reference. Panels e)-g) illustrate how the medians of the respective distributions in panels b)-d) vary with the tail volume selected in panel a). The circles mark the medians for the segmented distributions shown in panels b)-d) for the case of $10\%$ tail volume. For panels e) and f), the fraction of each quantity ($\epsilon$, $\chi$) contained within each tail relative to the entire domain is also indicated. The dashed diagonal lines mark what would be expected for a uniform distribution of each quantity throughout the domain.}
	\label{fig:Fig2}
\end{figure}

Motivated by the flow segmentation of \cite{portwood2016robust} in terms of the local fraction of overturning $\partial \rho / \partial z > 0$, we first consider how the pointwise distributions of $\epsilon$, $\chi$ and $\Gamma = \chi/ \epsilon$ depend on the magnitude of $\partial \rho / \partial z$, for both statically stable and unstable points, as shown in Figure \ref{fig:Fig2}. For illustration, in Figure \ref{fig:Fig2}a we split the distribution of $\partial \rho / \partial z$ into three regions: two tails containing 10$\%$ by volume of the most stable and unstable points (coloured blue and red, respectively), and the remaining 80$\%$ of the intermediate values (green). For such a division, we then consider the distributions of $\epsilon$, $\chi$ and $\Gamma$ within each region, as shown in Figures \ref{fig:Fig2}b-d. \mc{Although the distribution characterizing the bulk of the domain (green) is centered around the canonical flux coefficient $\Gamma = 0.2$ (see Figure \ref{fig:Fig2}d),} such points contain the lowest $\chi$ (Figure \ref{fig:Fig2}c) and are thus not of primary importance for the total mixing arising within the computational domain. Instead, it is the extreme tails of the $\partial \rho / \partial z$ distribution that must be considered, containing the most significant values of $\chi$. While both the blue and red tails contain elevated but similar distributions of $\epsilon$, they may be distinguished by their asymmetry in $\chi$; the stable tail (blue) contains disproportionately elevated $\chi$ as compared to $\epsilon$, and therefore some of the highest values of $\Gamma$ within the domain.

In Figures \ref{fig:Fig2}e-g, we then analyze how the medians of the $\epsilon$, $\chi$ and $\Gamma$ distributions change as a function of the volume contained within the blue and red tails, and additionally plot the relative contributions of these tails to the domain total. Comparing the right panels of Figures \ref{fig:Fig2}e and \ref{fig:Fig2}f reveals that $\chi$ is far more dominated by extreme events than $\epsilon$, in agreement with the analysis of oceanographic data by \cite{couchman2021data}. For instance, when each tail contains $10\%$ volume, the stable (blue) and unstable (red) tails each contain approximately $20\%$ of the total $\epsilon$ in the domain, but $45\%$ and $30\%$ of the total $\chi$, respectively. Furthermore, while the contributions to $\epsilon$ from both tails is roughly equal, the contribution to $\chi$ from the stable tail is always roughly $50\%$ greater than for the unstable tail. While $\Gamma=0.2$ may thus be a suitable approximation for the bulk of the domain, it may here not be relied upon for capturing the most extreme events in $\chi$, which dominate the bulk mixing statistics.

Additionally, the statistics in Figure \ref{fig:Fig2} suggest that local instability may not be a sufficient indicator for mixing, given the significance of the blue stable tail. However, we note that such a conclusion cannot definitively be drawn from the \emph{pointwise} distributions of $\partial \rho / \partial z$, as such a distribution provides no information about the extended spatial environment around each point. For example, in regions of fully-developed turbulence that might emerge after the collapse of a shear-induced billow, there is likely a random mixture of neighbouring unstable and stable points in close proximity (roughly a $50\%$ mixture as identified by \cite{portwood2016robust} in their most turbulent patches), and so points within the red and blue tails of Figure  \ref{fig:Fig2}a could be direct neighbours in space. Therefore, in $\S$\ref{sec:4_MixingStructures}, we extend our pointwise analysis by identifying spatially extended and coherent stable regions, which appear to take the form of `interfaces' with enhanced density gradients. We then assess the significance of these non-overturning structures to the overall mixing statistics.

\begin{figure}
	\captionsetup{width=1\linewidth}
	\centerline{\includegraphics[width=1\textwidth]{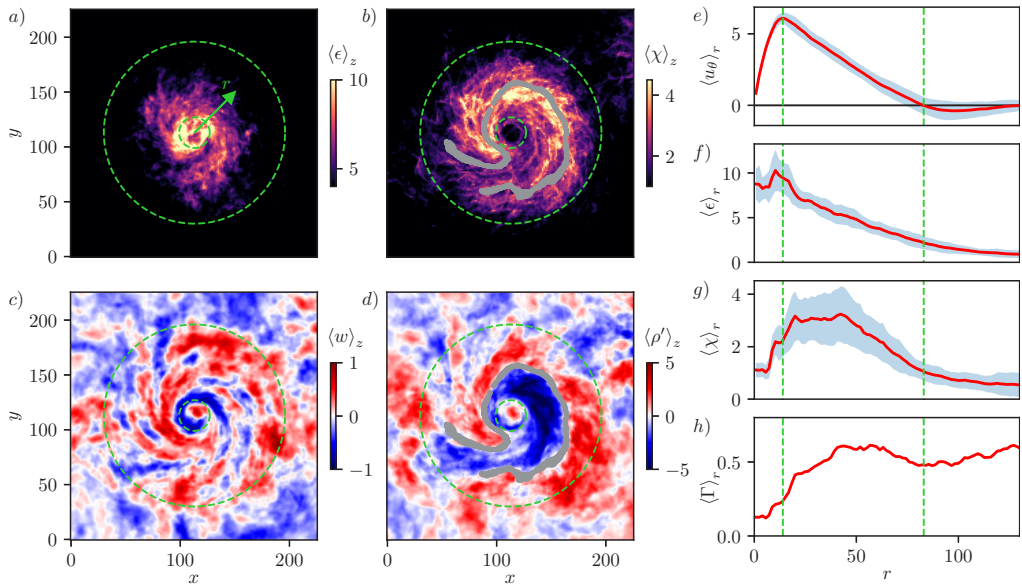}}
	\caption{Mixing patterns within the vortex. Vertical averages of the a) dissipation rate of kinetic energy $\epsilon$, b) dissipation rate of scalar variance $\chi$, c) vertical velocity $w$, and d) density perturbations $\rho'$, for the vortex region delineated in Figure \ref{fig:Fig1}a. The outer green circle in panels a)-d) coincides with the black circle in Figure \ref{fig:Fig1}a. The gray curve in panel d) delineates the sharp lateral gradient separating regions of positive and negative $\rho'$ and is found to be correlated with the spiral distribution of $\chi$ observed in panel b). Radial averages of e) the azimuthal velocity $u_\theta$, f) $\epsilon$, g) $\chi$ and h) $\Gamma$, as a function of the distance $r$ from the center of the vortex, with shading denoting the standard deviation around the radial mean. \mc{In panel h), $\left\langle \Gamma\right\rangle _{r}=\left\langle \chi\right\rangle _{r}/\left\langle \epsilon\right\rangle _{r}$ is the ratio of the red lines in panels f) and g).} The vertical green dashed lines in panels e)-h) mark the radial locations of the maximum and first zero of the radially-averaged azimuthal velocity, and correspond to the radii of the green dashed circles in panels a)-d).}
	\label{fig:Fig3}
\end{figure}

\section{Extended mixing structures} \label{sec:4_MixingStructures}
We now consider coherent spatial distributions of the microstructure quantities $\epsilon$ and $\chi$, and their relation to the large-scale flow patterns observed in Figure \ref{fig:Fig1}, by focusing on mixing structures arising both within and outside of the vortex. \mc{We first perform a closer examination of mixing within the vortex, as shown in Figure \ref{fig:Fig3}. Vertical averages of $\epsilon$, $\chi$, $w$ and $\rho'$ are plotted in Figures \ref{fig:Fig3}a-d, respectively, highlighting clear differences in the spatial distributions of $\epsilon$ and $\chi$. Such differences are further illustrated in Figures \ref{fig:Fig3}e-h, which show the respective radial distributions of the azimuthal velocity $u_\theta$, $\epsilon$, $\chi$ and $\Gamma$ with respect to the vortex core. These radial distributions illustrate that} the inner section of the vortex, characterized by roughly rigid-body rotation, is well-mixed and contains the largest values of $\epsilon$ despite having minimal scalar diffusion rates $\chi$. This observation is consistent with the density field shown in Figure \ref{fig:Fig1}d, where initially horizontal contours of constant density (green lines) are strongly deflected toward the vertical before reaching the center of the vortex, resulting in a vertically-extended core of roughly constant density (seen predominantly in the vertical interval $25\lesssim z\lesssim175$). Conversely, the majority of $\chi$ is found outside the core at radii where the angular velocity begins to decay, and is distributed in a roughly spiral pattern (Figure \ref{fig:Fig3}b). Examination of the vertically-averaged perturbed density field $\rho'$ (Figure \ref{fig:Fig3}d) reveals the presence of a strong lateral density gradient, induced by the alternating upwelling of dense fluid and downwelling of lighter fluid within the vortex as a function of $r$. Superimposing the position of this lateral gradient (gray) onto the distribution of $\chi$ in Figure \ref{fig:Fig3}b reveals that this gradient is strongly correlated with the spiral distribution of the most intense $\chi$.
While the vortex was identified by \cite{portwood2016robust} to be a patch of vigorous turbulence with elevated $\epsilon$ due to its generation of significant local vertical overturning, our analysis suggests that much of the mixing within the vortex, as quantified by $\chi$, instead results from diffusion across a strong lateral gradient in the perturbed density field.

\begin{figure}
	\captionsetup{width=1\linewidth}
	\centerline{\includegraphics[width=1\textwidth]{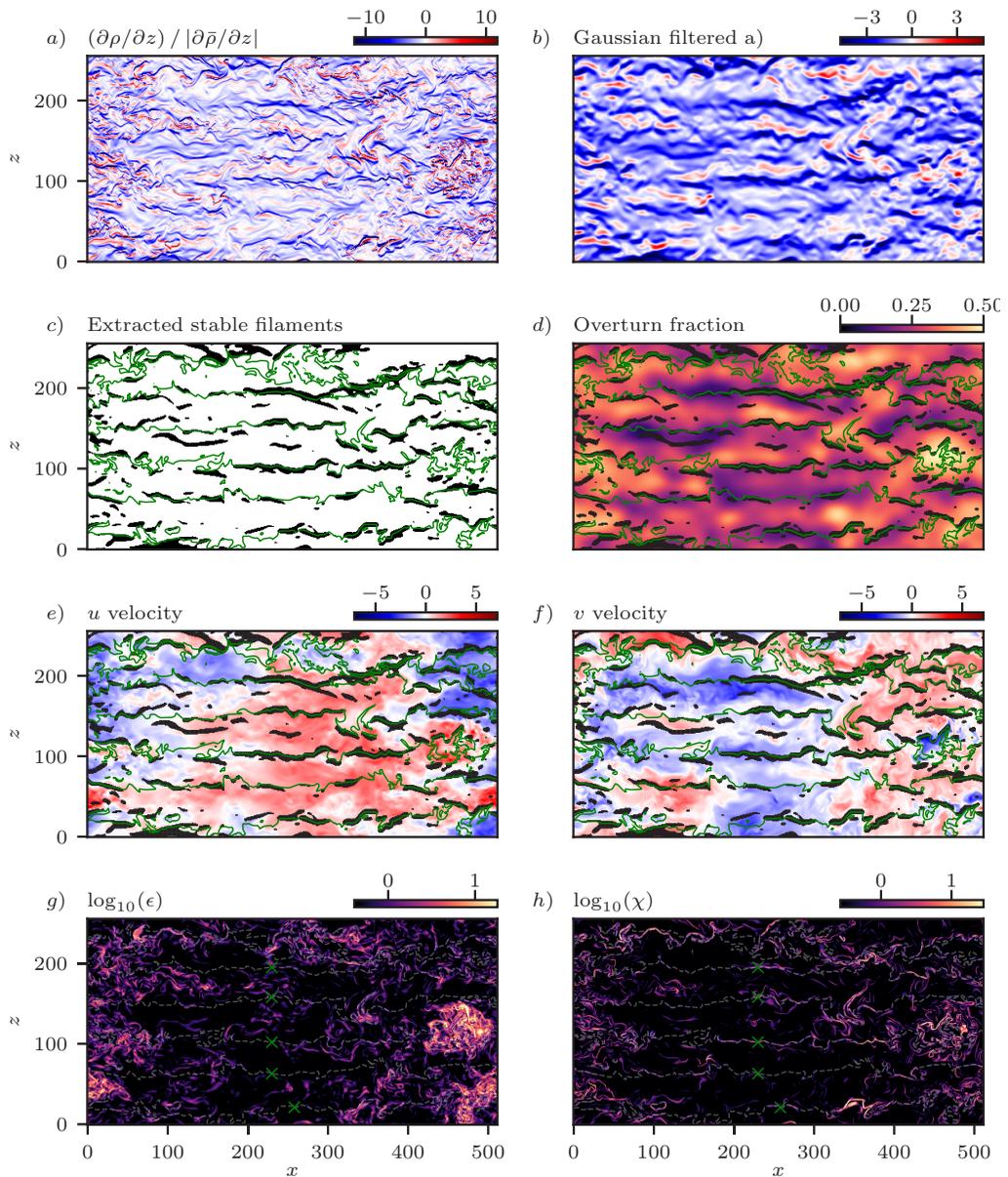}}
	\caption{Characteristics of the layered flow outside the vortex. Vertical slices at $y=200$ in Figure \ref{fig:Fig1}a of: a) the density field $\partial \rho / \partial z$ normalized by the magnitude of the imposed background gradient; b) the result of applying a Gaussian filter with a standard deviation of two gridpoints ($\approx 6L_K$) to the density field in panel a); c) extracted stable filaments from panel b) obtained by retaining points in the bottom (most-stable) $15\%$ of the filtered density distribution; d) the local fraction of unstable overturned points computed with a filter size corresponding to the Taylor length ($L_T \approx 25L_K$), following the method of \cite{portwood2016robust}; e-f) the $x$ and $y$ horizontal components of velocity, respectively; \mcc{and g-h) the dissipation rates of kinetic energy $\epsilon$ and scalar variance $\chi$, respectively.} The stable filaments from panel c) are overlaid on panels d)-f) for reference. The contours coloured green in panels c-f) \mcc{and white in panels g-h)} correspond to those plotted in Figures \ref{fig:Fig1}b,d, marking the stable interfaces separating relatively well-mixed density layers. \mcc{Green crosses in panels g-h) indicate examples of locations where $\chi$ is locally high, due to the presence of a stable density interface, without a corresponding increase in local $\epsilon$.}}
	\label{fig:Fig4}
\end{figure}

Outside of the vortex, the vertical homogeneity of the velocity and density fields collapses, forming a vertically layered structure. In Figure \ref{fig:Fig4}, we consider a vertical $(x,z)$ slice of the domain at position $y=200$ in Figure \ref{fig:Fig1}a, in order to understand how this large-scale layering pattern gives rise to mixing at the microscale. Motivated by the significance of the stable tail (blue) in the pointwise distribution of $\partial \rho / \partial z$ in Figure \ref{fig:Fig2}a, and the observation of horizontally-extended stable filaments of $\partial \rho / \partial z$ in Figure \ref{fig:Fig4}a, we examine whether such structures contribute substantially to mixing in the layered flow surrounding the vortex. To isolate these stable filaments, we apply a Gaussian filter to the density field with standard deviation $\sigma \approx 6 L_K$ (corresponding to 2 grid points in Figure \ref{fig:Fig4}), where $L_K$ denotes the Kolmogorov length scale. The intent of such a filter is to isolate spatially-coherent stable structures from patchy overturning regions that would contain a random assortment of stable and unstable neighbouring points. We note that our filter length is on the order of $10 L_K$ as suggested by \cite{kuo1971experiments} for removing internal intermittency. Further, it is significantly finer than the Taylor length $L_T \approx 25 L_K$, which was the smallest filter size considered by \cite{portwood2016robust} in their identification of `intermittent layers', allowing us to examine the importance of finer-scale structures within the flow. 

Having filtered the density field (Figure \ref{fig:Fig4}b), we then extract the most stable density structures by considering points in the bottom (most stable) $q$ percent of the filtered $\partial \rho / \partial z$ distribution. For illustration, we here extract structures comprised of the most stable $q=15\%$ of points (Figure \ref{fig:Fig4}c), and in Appendix \ref{AppA} demonstrate the effect of changing this percentage. The green contours from Figures \ref{fig:Fig1}b,d are overlaid on Figure \ref{fig:Fig4}c, demonstrating that the extracted filaments correspond to segments of the sharp interfaces separating relatively well-mixed layers in the density field. Importantly, Figure \ref{fig:Fig4}d highlights that the \mcc{concentration of locally-overturned points} (the segmentation indicator used by \citet{portwood2016robust}) is greatest in the regions between these stable interfaces. In Figures \ref{fig:Fig4}e,f, we again highlight that these stable interfaces are also roughly correlated with regions of high vertical shear in the layered velocity field, as are generated by the vortex. \mcc{Corresponding slices of the dissipation rates of kinetic energy $\epsilon$ and scalar variance $\chi$ are shown in Figures \ref{fig:Fig4}g,h, respectively. The spatial distribution of $\epsilon$ is seen to be much more diffuse than that of $\chi$, with extreme values of $\chi$ being primarily concentrated within thin filamentary structures such as those identified in Figure \ref{fig:Fig4}c. Crucially, there are many examples of locations in the flow (see green crosses, Figures \ref{fig:Fig4}g,h) where the stable interfaces identified in Figure \ref{fig:Fig4}c contain highly-elevated local signatures of $\chi$ without a proportional local increase in $\epsilon$. Such an observation thus raises the question as to whether these stable interfaces contribute significantly to the total mixing within the domain, in addition to the mixing occurring in more conventionally-studied isotropic overturning regions (such as the large overturn located in the vicinity of $(x,y)=(450,125)$ in Figure \ref{fig:Fig4}).}

\begin{figure}
	\captionsetup{width=1\linewidth}
	\centerline{\includegraphics[width=1\textwidth]{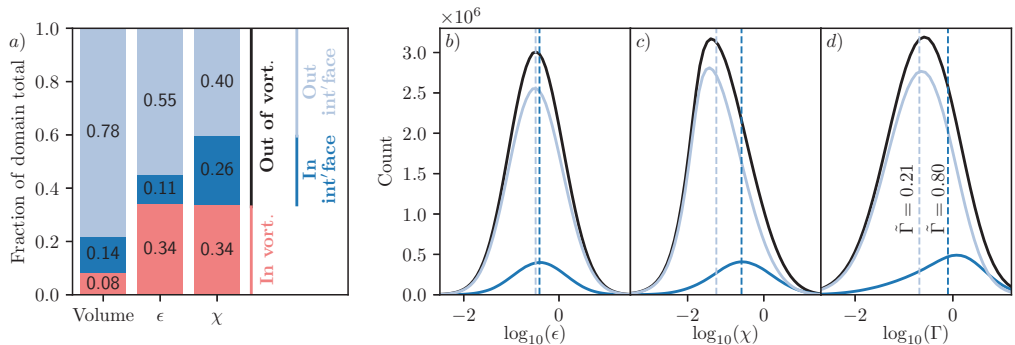}}
	\caption{Mixing contributions of the vortex and stable interfaces. a) Fractional contributions to the domain total of the dissipation rates of kinetic energy $\epsilon$ and scalar variance $\chi$, from within the vortex (red), stable interfaces outside the vortex (dark blue), as illustrated in Figure \ref{fig:Fig4}c, and the rest of the domain outside both the vortex and interfaces (light blue). Histograms of b) $\epsilon$, c) $\chi$ and d) $\Gamma$ outside  the vortex (black), further split according to whether points are contained within a stable interface (dark blue) or not (light blue). Dashed lines indicate median values of the respective distributions.}
	\label{fig:Fig5}
\end{figure}

We address the question \mcc{of whether the identified stable filaments contribute substantially to the total mixing occuring within the domain} in Figure \ref{fig:Fig5}, where we consider the relative contributions of both the vortex and the isolated stable interfaces to the domain totals of $\epsilon$ and $\chi$. In agreement with \citet{portwood2016robust}, despite occupying less than $10\%$ of the domain volume, the vortex contributes approximately a third of the entire domain's $\epsilon$ and $\chi$ (red bars, Figure \ref{fig:Fig5}a). Outside of the vortex, however, it is the stable interfaces that play a key role in the overall mixing, contributing
\mc{\begin{equation}
\frac{\textrm{(\% in interface) \ensuremath{\cap} (\% outside vortex)}}{(\textrm{\% outside vortex) }}=\frac{26\%}{26\%+40\%}=39\%
\end{equation}}
of the total $\chi$ outside the vortex, despite appearing unremarkable based on their much smaller contribution to $\epsilon$ (\mc{$11\%/\left(11\%+55\%\right)=17\%$}). Figure \ref{fig:Fig4}d thus highlights a key conclusion of this study: while \mcc{the concentration of overturned points} is most prevalent within the well-mixed density layers, relatively thin stable interfaces between such relatively deep layers, which are also correlated with high vertical shear, yield a crucial component of the bulk scalar mixing rate $\chi$. In particular, Figures \ref{fig:Fig5}b,c highlight that while these interfaces may be strongly distinguished by their distributions of $\chi$, \mc{where the median values differ by almost an order of magnitude}, they are virtually indistinguishable based on their distributions of $\epsilon$. This mismatch between the spatial distributions of $\epsilon$ and $\chi$ results in significantly elevated $\Gamma$ within the interfaces, well above the canonical value $\Gamma = 0.2$ (Figure \ref{fig:Fig5}d). It thus appears crucial to consider the independent information provided by the distributions of $\epsilon$ and $\chi$ within a domain when quantifying mixing, particularly for identifying the locations of the most extreme scalar mixing events. 

\section{Discussion} \label{sec:5_Discussion}
We have considered local correlations between the vertical density gradient $\partial \rho / \partial z$ and the dissipation rates of kinetic energy $\epsilon$ and scalar variance $\chi$ in order to characterize the spatial distributions of mixing within a forced direct numerical simulation of density-stratified turbulence. The forcing scheme is found to generate a vertically-aligned vortex within the domain, largely explaining the concentrated `patch' region of vigorous turbulence reported by \cite{portwood2016robust}. Outside of the vortex, the flow is characterized by a layered density profile, with thin, highly stable interfaces separating relatively well-mixed layers. While a mixing analysis based solely on the identification of local overturning would deem the well-mixed layers to be of primary importance, as in the identification of `intermittent layers' with elevated $\epsilon$ by \cite{portwood2016robust}, we have demonstrated that a significant fraction of $\chi$ is localized to the edges of such layers, within the \emph{stable} intervening interfaces. Notably, these interfaces appear unremarkable if looking at $\epsilon$ alone (see Figure \ref{fig:Fig5}b), emphasizing the importance of $\chi$ as an independent indicator of \mcc{local} mixing. \mcc{A number of other studies have also highlighted that significant mixing rates may be found in regions devoid of local overturning, emphasizing the importance of considering other mixing mechanisms present within stratified flows. For instance, by considering a different class of forced direct numerical simulations to those analyzed here, \cite{basak2006dynamics} demonstrated that horizontal shear is able to generate a complex pattern of vorticies which efficiently mix the density field without local overturning. A striking experimental demonstration of dye being transported across stationary, highly-stable density interfaces has been demonstrated by \cite{oglethorpe2013spontaneous}, where a ``scouring'' rather than overturning dynamic generates the mixing.} As the flux coefficient $\Gamma = \chi / \epsilon$ has been found to strongly depend on the time history of a turbulent event \citep{mashayeketal21}, it would be instructive to now consider the time evolution and formation of the stable interfaces identified in our study, characterized by strongly elevated $\Gamma$. For instance, as the density interfaces are correlated with regions of high vertical shear, it is conceivable that they might be remnants of the previous collapse of shear-induced billows that are now only visible in signals of $\chi$ but not $\epsilon$, as coined `fossil turbulence' by \cite{nasmyth1970oceanic}.

Our findings have two \mc{potential} implications for the parameterization of ocean mixing. Firstly, our analysis highlights the importance of adequately sampling rare, yet extreme mixing events in a turbulent flow, \mc{as was also recently discussed by \citet{cael2021log}}. In agreement with the analysis of oceanographic data by \cite{couchman2021data}, Figure \ref{fig:Fig2}d demonstrates that although the majority of the domain indeed appears to be well characterized by the canonical flux coefficient $\Gamma = 0.2$, significantly elevated $\Gamma$ is associated with the most extreme events in $\chi$, events that are not reflected by a corresponding local increase in $\epsilon$. Given the current relative sparsity of measurements within the ocean, mixing parameterizations may thus be biased toward the most commonly measured events, which are not necessarily the most significant. \mc{Secondly, even with perfect sampling, different proxies for mixing are likely to yield contrasting predictions for the amount and spatial distribution of mixing within the highly-anisotropic layered flow considered here. For example, if measurements of $\chi$ were not available, the stable filaments at the edges of the overturning layers (Figure \ref{fig:Fig4}d) would appear unremarkable, as they appear locally quiescent based on their density gradient and are not correlated with any discernible increase in $\epsilon$. Further, given the strong spatial variability of $\Gamma$ within the vertically-layered flow (see Figure \ref{fig:Fig5}d), it is unclear what value of $\Gamma$ should be used in the method of \cite{osborn1980estimates} if trying to infer a flux from values of $\epsilon$ measured directly by a microstructure profiler or derived from a Thorpe overturning analysis.}

\mc{As discussed by \cite{caulfield2021layering}, an accurate parameterization of the flux coefficient $\Gamma$ is likely to depend on multiple dimensionless groups characterizing the underlying flow, such as the buoyancy Reynolds number $Re_b$, Froude number $Fr$, and Prandtl number $Pr$. For instance, DNS studies have demonstrated that bulk-averages of $\Gamma$ decrease with increasing $Pr$ \citep{salehipour2015turbulent} and decreasing $Fr$ \citep{maffioli2016mixing}. A promising future direction of inquiry would be to try and rationalize such variations in $\Gamma$ in terms of differences in the prevalence and structure of smaller-scale extreme events within the flow, such as analyzing changes in the morphology of the stable filaments considered here. It would also be instructive to extend our analysis to simulations of decaying turbulence which also develop layered structures \citep{de2019effects}, in order to establish whether the spatial distribution of mixing events observed here changes significantly in forced versus unforced scenarios.}

Finally, following \cite{portwood2016robust} and typical oceanographic measurements, we have here primarily relied upon the local density gradient $\partial \rho / \partial z$ to inform our analysis of spatial mixing patterns. However, there are likely more optimal flow variables, or linear combinations thereof, that could lead to a more robust segmentation of the turbulent domain into distinct regimes. For example, one could imagine constructing more insightful indicator functions of mixing from components of the velocity gradient tensor $\partial u_i / \partial x_j$, as suggested by \cite{kops2019unsupervised}. Applying data-driven techniques, such as unsupervised clustering or dimensionality reduction, to the wealth of observational, experimental and numerical stratified turbulence data currently available has the potential to discover automatically optimal mixing indicators free of human bias. Such an analysis would hopefully further our understanding of the dominant mixing mechanisms present in different flow regimes, along with their prevalence, guiding the search for a more universal and accurate mixing parameterization.

\backsection[Acknowledgements]{
This research used resources of the Oak Ridge Leadership Computing Facility at the Oak Ridge National Laboratory, which is supported by the Office of Science of the U.S. Department of Energy under Contract No. DE-AC05-00OR22725.  SdeBK was supported under U.S. Office of Naval Research Grant number N00014-19-1-2152. For the purpose of open access, the authors have applied a Creative Commons Attribution (CC BY) licence to any Author Accepted Manuscript version arising from this submission.}


\backsection[Declaration of interests]{The authors report no conflict of interest.}


\appendix
\section{Thresholding of stable filaments}\label{AppA}
The stable filaments (black) plotted in Figure \ref{fig:Fig4}c were extracted by identifying points within the bottom (most stable) $q=15\%$, by volume, of the Gaussian-filtered distribution of the density gradient $\partial \rho / \partial z$ (Figure \ref{fig:Fig4}b). We here briefly consider how changing this thresholding percentage $q$ influences the characteristics of the extracted stable structures.

\begin{figure}
	\captionsetup{width=1\linewidth}
	\centerline{\includegraphics[width=1\textwidth]{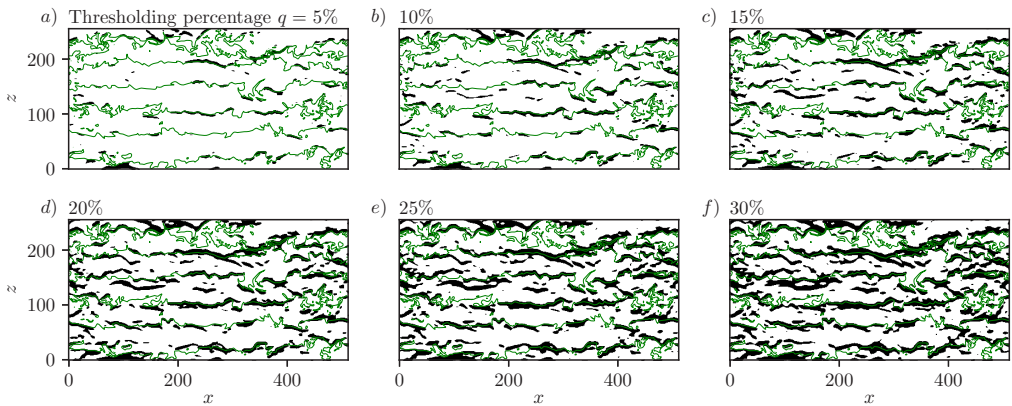}}
	\caption{Extracted stable filaments as a function of the thresholding percentage $q$. Filaments (black) are identified as the union of points within the bottom (most stable) $q\%$ of the filtered density distribution (see discussion in $\S$\ref{sec:4_MixingStructures}). The filaments detected using values of $q$ between $5\%$ and $30\%$ are shown in panels a)-f), respectively. Panel c) corresponds to Figure \ref{fig:Fig4}c.}
	\label{fig:Fig6A_App}
\end{figure}

In Figure \ref{fig:Fig6A_App}, we plot the stable structures that are identified by varying the percentage $q$ from $5\%$ to $30\%$. As $q$ is increased, meaning that more points in the stable tail of the filtered $\partial \rho / \partial z$ distribution are considered, the identified stable structures are found to grow primarily in the horizontal direction, tracing out more of the stable interfaces identified by the green contours from Figures \ref{fig:Fig1}b,d. Figure \ref{fig:Fig6A_App} thus highlights that the magnitude of the vertical density gradient along such stable contours is not uniform, with certain segments having stronger gradients (as identified by using a smaller $q$) and thus being characterized by a larger local $\chi$.

\begin{figure}
	\captionsetup{width=1\linewidth}
   	\centerline{\includegraphics[width=1\textwidth]{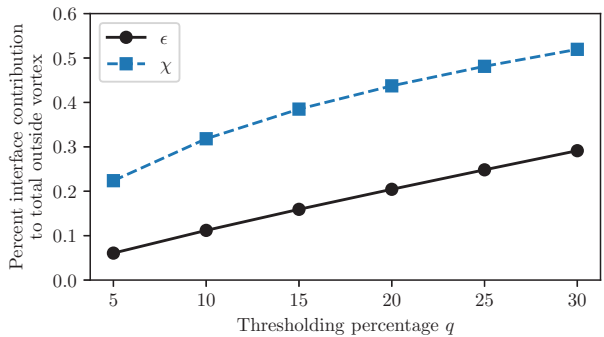}}
	\caption{The percent contributions of $\epsilon$ and $\chi$ contained within the stable interfaces identified in Figure \ref{fig:Fig6A_App}, normalized by the domain totals outside of the vortex, as a function of the thresholding percentage $q$. The statistics at $q=15\%$ correspond to those presented in Figure \ref{fig:Fig5}a, noting that in Figure \ref{fig:Fig5}a the contributions are normalized by the total domain including the vortex.}
	\label{fig:Fig6B_App}
\end{figure}

It is also natural to consider how the mixing statistics presented in Figure \ref{fig:Fig5}a depend on the thresholding percentage $q$. In Figure \ref{fig:Fig6B_App}, considering only the computational domain outside of the vortex, we plot the percent contribution of the extracted interfaces to $\epsilon$ and $\chi$, as a function of $q$. The points at $q=15\%$ correspond to the statistics presented in Figure \ref{fig:Fig5}a, noting that in Figure \ref{fig:Fig6B_App} the percent contributions are normalized by the domain total outside the vortex, and not the entire domain including the vortex as in Figure \ref{fig:Fig5}a. Figure \ref{fig:Fig6B_App} demonstrates that over a wide range of threshold percentages $q$, the identified stable filaments always contribute over twice the amount of $\chi$ as compared to $\epsilon$.  

\clearpage

\bibliographystyle{jfm}
\bibliography{LayersBib}

\end{document}